\documentclass[aps,prd,preprint,12pt,superscriptaddress,nofootinbib,a4paper]{revtex4-1}

\usepackage[utf8]{inputenc}
\usepackage{amsmath,amssymb}
\usepackage[separate-uncertainty=true]{siunitx}
\usepackage{graphicx}
\usepackage[usenames,dvipsnames]{xcolor}
\usepackage[caption=false]{subfig}
\usepackage{hyperref}
\usepackage[capitalize]{cleveref}
\usepackage{booktabs}
\usepackage{tabularx}
\usepackage{xspace}
\usepackage{color}
\usepackage[normalem]{ulem}
\usepackage{slashed}
\usepackage{bbold}
\usepackage{wasysym}
\usepackage{graphicx}
\usepackage{mathrsfs}

\allowdisplaybreaks

\graphicspath{{graphics/}}

\newcommand{\eqn}{equation}
\newcommand{\lb}{\left(}
\newcommand{\rb}{\right)}
\newcommand{\be}{\beta}
\newcommand{\al}{\alpha}

\newcommand{\GeV}{{\ensuremath\rm GeV}}
\newcommand{\TeV}{{\ensuremath\rm TeV}}
\newcommand{\lam}{\lambda}
\newcommand{\MeV}{{\ensuremath\rm MeV}}
\newcommand{\pb}{{\ensuremath\rm pb}}
\newcommand{\fb}{{\ensuremath\rm fb}}

\DeclareSIUnit{\pb}{pb}
\DeclareSIUnit{\fb}{fb}
\AtBeginDocument{
\heavyrulewidth=.08em
\lightrulewidth=.05em
\cmidrulewidth=.03em
\belowrulesep=.65ex
\belowbottomsep=0pt
\aboverulesep=.4ex
\abovetopsep=0pt
\cmidrulesep=\doublerulesep
\cmidrulekern=.5em
\defaultaddspace=.5em

\newcolumntype{C}{>{\centering\arraybackslash}X}
\newcolumntype{b}{C}
\newcolumntype{s}{>{\hsize=.6\hsize}C}
\newcolumntype{R}{>{\raggedleft\arraybackslash}X}
}

\begin{document}
\bibliographystyle{hunsrt}
\date{\today}
\rightline{RBI-ThPhys-2021-018}
\title{{\Large The THDMa revisited\\ - A preview -}}
\author{Tania Robens}
\email{trobens@irb.hr}
\affiliation{Ruder Boskovic Institute, Bijenicka cesta 54, 10000 Zagreb, Croatia}

\renewcommand{\abstractname}{\texorpdfstring{\vspace{0.5cm}}{} Abstract}

\begin{abstract}
    \vspace{0.5cm}
We here present preliminary results on a parameter scan of the THDMa, a new physics model that extends the scalar sector of the Standard Model by an additional doublet as well as a pseudoscalar singlet. In the gauge-eigenbasis, the additional pseudoscalar serves as a portal to the dark sector, with a fermionic dark matter candidate. This model is currently one of the standard benchmarks for the LHC experimental collaborations. We apply all current theoretical and experimental constraints and identify regions in the parameter space that might be interesting for an investigation at possible future $e^+e^-$ facilities.\\
{\sl Talk presented at the International Workshop on Future Linear Colliders (LCWS2021), 15-18 March 2021. C21-03-15.1}
\end{abstract}

\maketitle

\section{Introduction}

We present preliminary results for a parameter scan on the THDMa, a two Higgs doublet model (THDM) with an additional pseudoscalar that serves as a portal to dark matter \cite{Ipek:2014gua,No:2015xqa,Goncalves:2016iyg,Bauer:2017ota,Tunney:2017yfp}. It features 5 additional particles in the scalar sector, which we label $H,A,a,H^\pm$ in the mass eigenbasis, as well as a fermionic dark matter candidate $\chi$. The model contains 14 free parameters after electroweak symmetry breaking, out of which 2 are fixed by the measurement of the 125 \GeV~ scalar as well as electroweak precision measurements. The parameter space is subject to a large number of theoretical and experimental constraints, see e.g. \cite{Abe:2018bpo,Pani:2017qyd,Haisch:2018znb,Abe:2018emu,Haisch:2018hbm,Haisch:2018bby,Abe:2019wjw,Butterworth:2020vnb,Arcadi:2020gge} for more recent work on this model. We also impose bounds from experimental searches, where we found \cite{Chatrchyan:2013mxa,Aaboud:2017rzf,Aaboud:2018knk,Aaboud:2018eoy,Sirunyan:2019xls,Sirunyan:2020fwm,Aad:2020zxo,Aad:2020zmz,ATLAS-CONF-2021-006} to have impact on the models parameter space\footnote{Note that bounds are here applied successively, e.g. all theoretical constraints are applied prior to experimental bounds, etc.}. This work is a preview of more complete results \cite{tocome}, which will be published shortly. 
\section{The model}\label{sec:model}
The model discussed in this work has been introduced in references \cite{Ipek:2014gua,No:2015xqa,Goncalves:2016iyg,Bauer:2017ota,Tunney:2017yfp}, and we refer the reader to these works for a detailed discussion of the model setup. We here just list generic features for brevity, where we follow the nomenclature of \cite{Abe:2018bpo}.

The field content of the THDMa in the gauge eigenbasis consists of two scalar fields $H_{1,2}$ which transform as doublets under the $SU(2)\,\times\,U(1)$ gauge group, and an additional pseudoscalar $P$ transforming as a singlet, as well as a dark matter candidate $\chi$ which we choose to be fermionic. The THDM part of the potential is given by
\begin{align}
V_{\text{THDM}} =& \mu_1 H_1^\dagger H_1 + \mu_2 H_2^\dagger H_2 
+ \lambda_1 (H_1^\dagger H_1)^2 + 
\lambda_2 (H_2^\dagger H_2)^2
+ \lambda_3 (H_1^\dagger H_1)(H_2^\dagger H_2)\\
&+ \lambda_4 (H_1^\dagger H_2) (H_2^\dagger H_1)
 + \left[\mu_3 H_1^\dagger H_2 + {\lambda_5} (H_1^\dagger H_2)^2 + h.c.\right]
\end{align}
Fields are decomposed according to (see also e.g. \cite{Branco:2011iw})
\begin{\eqn}\label{eq:fieldpar}
H_i\,=\,\lb \begin{array}{c} \phi_i \\ \frac{1}{\sqrt{2}} \lb v_i+\rho_i+i\,\eta_i \rb \end{array} \rb
\end{\eqn}

where
\begin{\eqn}\label{eq:vevtb}
{v_1\,=\,v\,\cos\be,\;v_2\,=\,v\,\sin\be.}
\end{\eqn}

The scalar potential is 

\begin{align}\label{eq:vp}
V_P = 
  \frac{1}{2} m_P^2 P^2 + \lambda_{P_1} H_1^\dagger H_1 P^2 +
  \lambda_{P_2} H_2^\dagger H_2 P^2 +
  (\imath b_P H_1^\dagger H_2 P + h.c.)
\end{align}
Finally, the coupling between the visible and the dark sector is mediated via the interaction
\begin{\eqn*}
\mathcal{L}_\chi\,=\,-i\,y_\chi P\bar{\chi}\gamma_5\,\chi.
\end{\eqn*}
Couplings of the scalar sector to the fermionic sector arizes from
\begin{\eqn*}
\mathcal{L}_Y\,=\,-\sum_{i=1,2}\,\left\{ \bar{Q}\,Y^i_u\,\bar{H}_i\,u_R+\bar{Q}\,Y^i_d\,\bar{H}_i\,d_R + \bar{L}\,Y^i_\ell\,\bar{H}_i\,\ell_R\,+\,h.c.\right\},
\end{\eqn*}
where $Y^i_{u,d}$ denote the Yukawa matrices, $Q$ and $L$ are left-handed quark and lepton doublets, and $u_R,\,d_R,\,\ell_R$ label right-handed uptype, downtype, and leptonic gauge singlets. We impose an additional $Z_2$ symmetry on the model, under which the doublets transform as $H_1\,\rightarrow\,H_1,\,H_2\,\rightarrow\,-H_2$, in order to avoid contributions from flavour changing neutral currents. In this work, we concentrate on the case where $Y_1^u\,=\,Y_2^d\,=\,Y_2^\ell\,=\,0$, which corresponds to a type II classification of Yukawa couplings in the THDM notation.\\

After electroweak symmetry breaking, the model is characterized by in total 14 free parameters. The mixing in the THDM sector is customary described by the mixing angles $\al,\,\be$. Furthermore, $V_P$ introduces a mixing between the pseudoscalar part of the THDM and the new pseudoscalar $P$, with introduces an additional mixing angle $\theta$. We here choose
\begin{\eqn}\label{eq:pars}
v,\,m_h,\,m_A,\,m_H,\,m_{H^\pm},\,m_a,\,m_\chi,\,\cos\lb \be-\al\rb,\,\tan\be,\,\sin\theta,\,y_\chi,\,\lam_3,\,\lam_{P_1},\,\lam_{P_2}
\end{\eqn}
as free parameters, where $v$ and $m_h$ are fixed to be $\sim\,246\,\GeV$ and $\sim\,125\,\GeV$. The leftover 12 parameters can float freely, but are subject to theoretical and experimental constraints. 

\section{Theoretical and experimental constraints}\label{sec:const}
The models parameter space is subject to a list of theoretical and experimental constraints. Most of these have been discussed in previous publications, as e.g. \cite{Abe:2018bpo,Abe:2019wjw,Arcadi:2020gge}. For the results presented here, some bounds have been applied using private codes, others have been tested making use of publicly available tools such as SPheno \cite{Porod:2003um,Porod:2011nf}, Sarah \cite{Staub:2008uz,Staub:2009bi,Staub:2010jh,Staub:2012pb,Staub:2013tta}, HiggsBounds \cite{Bechtle:2008jh,Bechtle:2011sb,Bechtle:2013gu,Bechtle:2013wla,Bechtle:2015pma,hb,Bechtle:2020pkv}, HiggsSignals \cite{Stal:2013hwa,Bechtle:2013xfa,Bechtle:2014ewa,hb,Bechtle:2020uwn}, and MadDM \cite{Backovic:2013dpa,Backovic:2015cra,Ambrogi:2018jqj}.

The following bounds have been imposed:
\begin{itemize}
\item{}{\bf Perturbativity, perturbative unitarity, and positivity of the potential}, leading to inequalities involving the potential couplings.
\item{}{\bf Constraints from electroweak precision observables} via the oblique paramters $S,\,T,\,U$ \cite{Altarelli:1990zd,Peskin:1990zt,Peskin:1991sw}, comparing to values from the latest Gfitter results \cite{Baak:2014ora,gfitter,Haller:2018nnx}.
\item{}{\bf Constraints from $B\,\rightarrow\,X_s\,\gamma,\,B_s\,\rightarrow\,\mu^+\,\mu^-$, and $\Delta m_s$}. Details of the comparision will be given in \cite{tocome}. For $B\,\rightarrow\,X_s\,\gamma$, we use a two-dimensional fit function \cite{mm} that reflects the bounds derived in \cite{Misiak:2020vlo}. For the other variables, we compare (using \cite{Beneke:2019slt} for theory predictions and \cite{ATLAS-CONF-2020-049,Amhis:2019ckw} for experimental findings)
\begin{eqnarray*}
&&\lb B_s\,\rightarrow\,\mu^+\mu^-\rb^\text{th}\,=\,\lb 3.66\,\,\pm\,0.14 \rb\times\,10^{-9}\\
&&\lb B_s\,\rightarrow\,\mu^+\mu^-\rb^\text{comb, exp}\,=\,\lb 2.69^{+0.37}_{-0.35}\rb\times\,10^{-9},\\
&&\\
&&\lb \Delta m_s\rb^\text{th}\,=\,\lb 17.61\,\pm\,1.05\rb\, \text{ps}^{-1},\\
&&\lb \Delta m_s\rb^\text{exp}\,=\,\lb 17.757\,\pm\,0.020\,\pm\,0.007\rb\,\text{ps}^{-1}.
\end{eqnarray*}
The theoretical value $\lb \Delta m_s\rb^\text{th}$ has been derived using  \cite{Lenz:2010gu,Lenz:2011ti,ulidisc}, with input values \cite{Aoki:2019cca,Dowdall:2019bea,Zyla:2020zbs} $f_{B_s}\,=\,\lb 230.3\,\pm\,1.3\rb \MeV,\,\widehat{\mathcal{B}}_{B_s}\,=\,1.232\,\pm\,0.053 ,\,V_{cb}\,=\,\lb 42.2 \,\pm\,0.8 \rb\,\times\,10^{-3}$.
\item{}{\bf Upper limit on width}: We impose (using the limit on $\Gamma_h$ from \cite{Sirunyan:2019twz})
\begin{\eqn*}
\Gamma_{h,125}\,\leq\,9\,\MeV,\;\Gamma_i/ M_i\,\leq\,0.5,\;i\,\in\left\{H,A,H^\pm,a \right\}
\end{\eqn*}
\item{}{\bf Agreement with null-results from past and current searches and signal strength measurements}. These bounds have been implemented making use of  HiggsBounds and HiggsSignals, that use a factorized approach for signal rate predictions.
\item{}{\bf Dedicated LHC searches} The model has been searched for in various final states. We here include bounds from $\ell^+\ell^-\,+\,\text{MET}$ \cite{Sirunyan:2020fwm}, $h\,+\,\text{MET}$ \cite{ATLAS-CONF-2021-006}, $H^+\,\bar{t}b,\,H^+\,\rightarrow\,t\,\bar{b}$ \cite{Aad:2020kep,Aad:2021xzu}, $W\,t+\text{MET}$ \cite{Aad:2020zmz} and $t\,\bar{t}/ b \bar{b}+\,\text{MET}$ \cite{Aaboud:2017rzf,Aad:2020aob,Aad:2021hjy,Aad:2021jmg}. All of these, apart from \cite{Aaboud:2017rzf}, correspond to searches making use of full Run 2 data. All production cross sections have been calculated using  Madgraph5 \cite{Alwall:2011uj}, with the UFO model provided in \cite{Bauer:2017ota,ufo}. For details, we refer to the reader to \cite{tocome}.
\item{}{\bf Dark matter constrains} The calculation of dark matter relic density makes use of the tool MadDM. For direct detection, we implemented the analytic expressions presented in \cite{Ipek:2014gua}. We compare these values to limits from the Planck collaboration \cite{Aghanim:2018eyx}, and require that
\begin{\eqn}\label{eq:omup}
\Omega\,h^2\,\leq\,0.1224
\end{\eqn}
which corresponds to a 2 $\sigma$ limit. Direct detection bounds are compared to maximal cross section values $\sigma^\text{Xenon1T}_\text{max}\,\lb m_{\chi}\rb$  using XENON1T result \cite{Aprile:2018dbl}, which we implemented in terms of an approximation function\footnote{The numerical values have been obtained using the Phenodata database \cite{PhenoData}.}. Relic density constraints are rescaled using 
\begin{\eqn}\label{eq:resc}
\sigma_\text{max}\,\lb m_{\chi,i},\Omega_i \rb\,=\,\sigma^\text{Xenon1T}_\text{max}\,\lb m_{\chi}\rb\,\frac{0.1224}{\Omega_i},
\end{\eqn}
where $m_{\chi,i},\,\Omega_i$ refer to the dark matter and relic density of the specific parameter point $i$ tested here.
\end{itemize}

\section{Scan setup and results}\label{sec:scan}
For a detailed discussion of the scan setup and steps, we refer the reader to \cite{tocome}.

Our initial scan ranges are determined by a number of prescans to determine regions of parameter space that are highly populated:
\begin{eqnarray}\label{eq:ranges}
&&\sin\theta\,\in\,\left[-1;0.8\right];\;\cos\lb \be-\al\rb\,\in\left[ -0.08;0.1 \right];\,\tan\be\,\in\,\left[0.52;9\right], \nonumber\\
&&m_H\,\in\,\left[500;1000\right]\GeV,\,m_A\,\in\,\left[600; 1000 \right]\GeV,\,m_{H^\pm}\,\in\left[800;1000\right]\GeV, \nonumber\\
&&m_a\,\in\,\left[0;m_A\right],\,m_\chi\,\in\,\left[0,m_a\right], \nonumber\\
&&y_\chi\,\in\,\left[-\pi;\pi\right],\,\lambda_{P_1}\,\in\,[0;10],\,\lambda_{P_2}\,\in\,\left[0;4\,\pi\right],\,\lambda_3\,\in\left[-2;4\,\pi\right].
\end{eqnarray}
The fit for $B\,\rightarrow\,X_s\,\gamma$ implies a lower bound on $m_{H^\pm}$ of $\sim\,800\,\GeV$ that is directly implemented in the scan setup.
The values of $m_h\,=\,125\,\GeV$ and $v\,=\,246\,\GeV$ are set according to measurements of the Higgs boson mass as well as electroweak precision measurements. Values outside the above regions are not forbidden; we chose the scan ranges to optimize parameter point generation performance.
\subsection{Scan results}
In the following, we discuss the resulting constraints on the parameter space of the THDMa. Note that not all bounds discussed above lead to a direct limit in a two-dimensional parameter plane. In particular:

\begin{itemize}
\item{}B-physics constraints set a lower bound on $\tan\be$ as a function of $m_{H^\pm}$, see figure \ref{fig:brsg}; in general, $\tan\be\,>\,1$.
\begin{center}
\begin{figure}[tb]
\begin{center}
\includegraphics[width=0.6\textwidth]{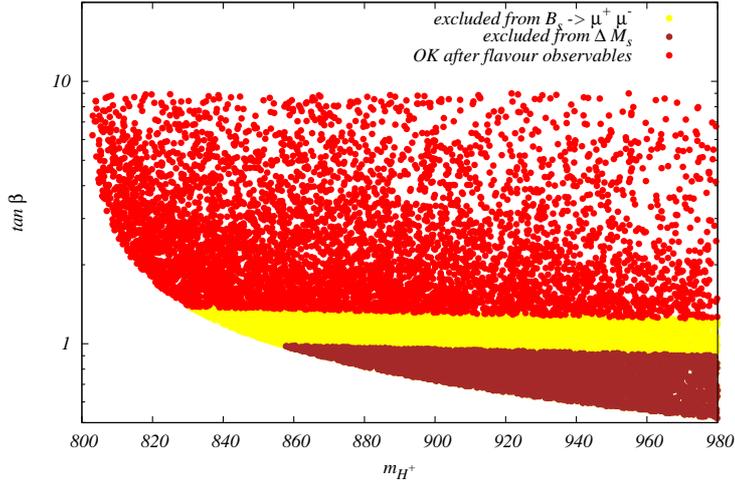}
\caption{\label{fig:brsg} Exclusion in the $\lb m_{H^\pm},\tan\be\rb$ plane after applying flavour constraints. The lower bound for $m_{H^\pm}\,\leq\,850\,\GeV$ is set by the bound on $B\,\rightarrow\,X_s\,\gamma$.}
\end{center}
\end{figure}
\end{center}
\item{}Oblique parameters reduce the allowed mass differences, see figures \ref{fig:stuexcl} and \ref{fig:stucomp}. The latter also shows comparison with bounds in the THMD decoupling limit. We see that allowing for an admixture with the second pseudoscalar in the THMDa enlarges the allowed parameter space.

\begin{center}
\begin{figure}[tbh]
\begin{center}
\includegraphics[width=0.6\textwidth]{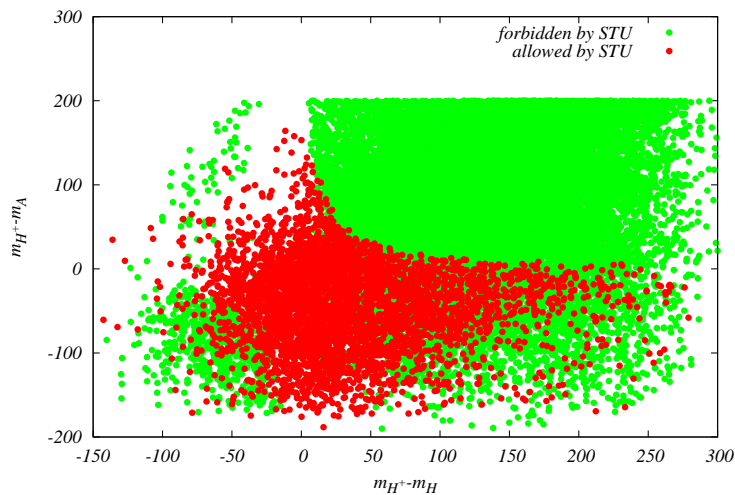}
\caption{\label{fig:stuexcl} Exclusions in the $\lb m_{H^\pm}-m_H,m_{H^\pm}-m_A\rb$ plane from oblique parameters. We see that regions where both displayed mass differences are large are excluded by the oblique parameters.  }
\end{center}
\end{figure}
\end{center}

\begin{center}
\begin{figure}[tbh]
\begin{center}
\includegraphics[width=0.45\textwidth]{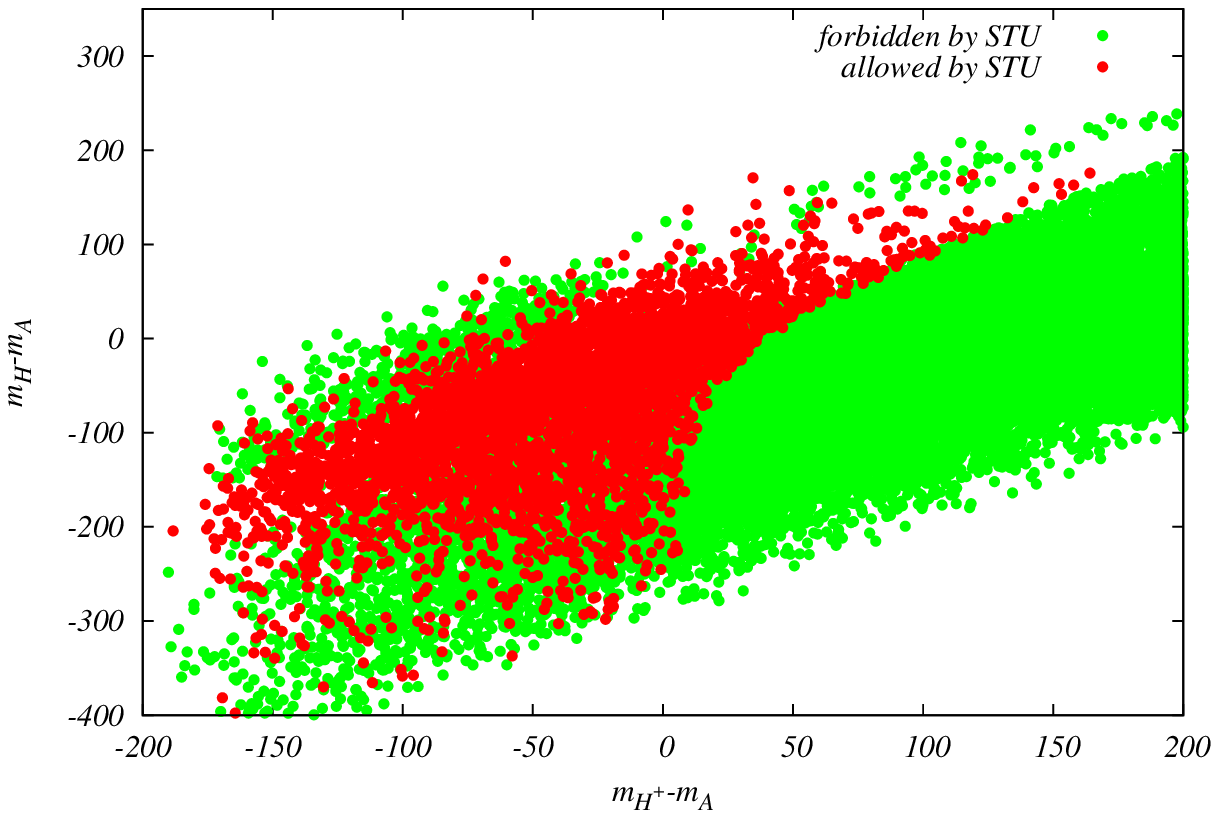}
\includegraphics[width=0.45\textwidth]{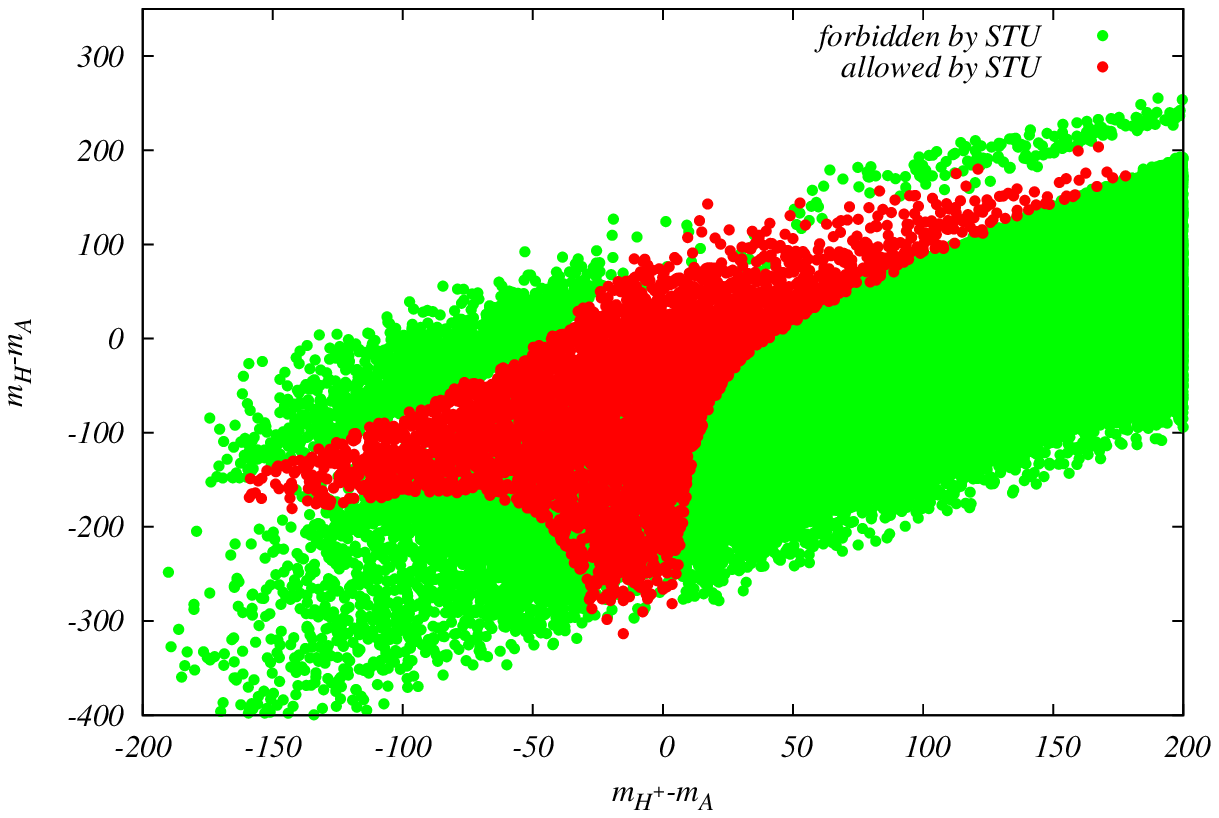}
\caption{\label{fig:stucomp} Exclusions in the $\lb m_{H^\pm}-m_A,m_{H}-m_A\rb$ plane in the THDMa {\sl (left)} and THDM {\sl (right)} from oblique parameters; for the latter, $\sin\theta\,=\,\lam_{P_1}\,=\,\lam_{P_2}\,=\,0$. The admixture of $a$ releases the bounds, as expected.}
\end{center}
\end{figure}
\end{center}
\item{}Signal strength measurements reduced the available parameter space for $\cos\lb\be-\al\rb$, such that now $\cos\lb\be-\al\rb\,\in\left[ -0.04;0.04\right]$, see figure \ref{fig:hbhs}. Some parameter points are also excluded by direct searches, $H/ a\,\rightarrow\,\tau\,\tau$ \cite{Aad:2020zxo}, $H\,\rightarrow\,h_{125}h_{125}$ \cite{Aaboud:2018knk} and $H\,\rightarrow\,a\, Z$ \cite{Aaboud:2018eoy,Sirunyan:2019xls} searches. Values of $\cos\lb \be-\al\rb\,>\,0.04$ and $\tan\be\,\gtrsim\,5$ are excluded from $h_{125}\,\rightarrow\,Z\,Z$ \cite{Chatrchyan:2013mxa}.
\begin{center}
\begin{figure}[tbh]
\begin{center}
\includegraphics[width=0.6\textwidth]{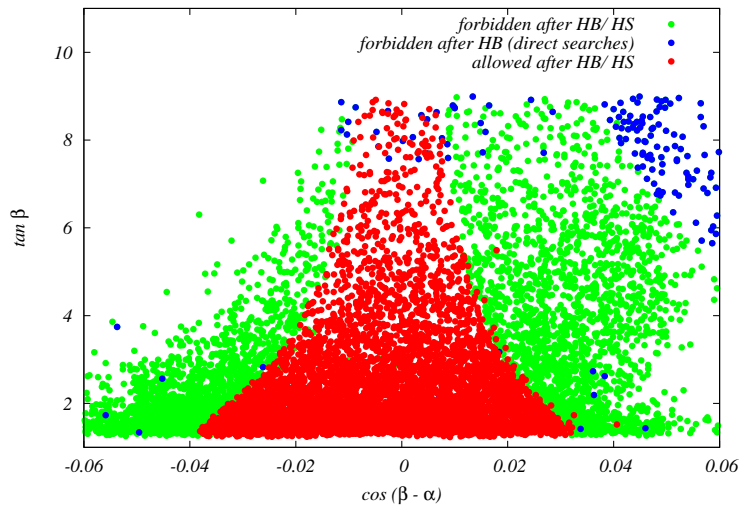}
\caption{\label{fig:hbhs} Exclusion in the $\cos(\be-\al),\tan\be$ plane after HiggsBounds (HB) and HiggsSignals (HS). }
\end{center}
\end{figure}
\end{center}
\item{}Relic density reduced the available parameter space to regions where $m_a\,-\,2\,m_\chi\,\in\left[-100;300 \right]\,\GeV$. Dominant annihilation channels are $\chi \bar{\chi}\,\rightarrow\,b\,\bar{b}$ and $\chi \bar{\chi}\,\rightarrow\,t\,\bar{t}$, where the latter channel opens up above the $t\,\bar{t}$ threshold where $m_a\,\gtrsim\,2\,m_t$. Results are displayed in figure \ref{fig:om}.
\begin{center}
\begin{figure}[tbh]
\begin{center}
\includegraphics[width=0.45\textwidth]{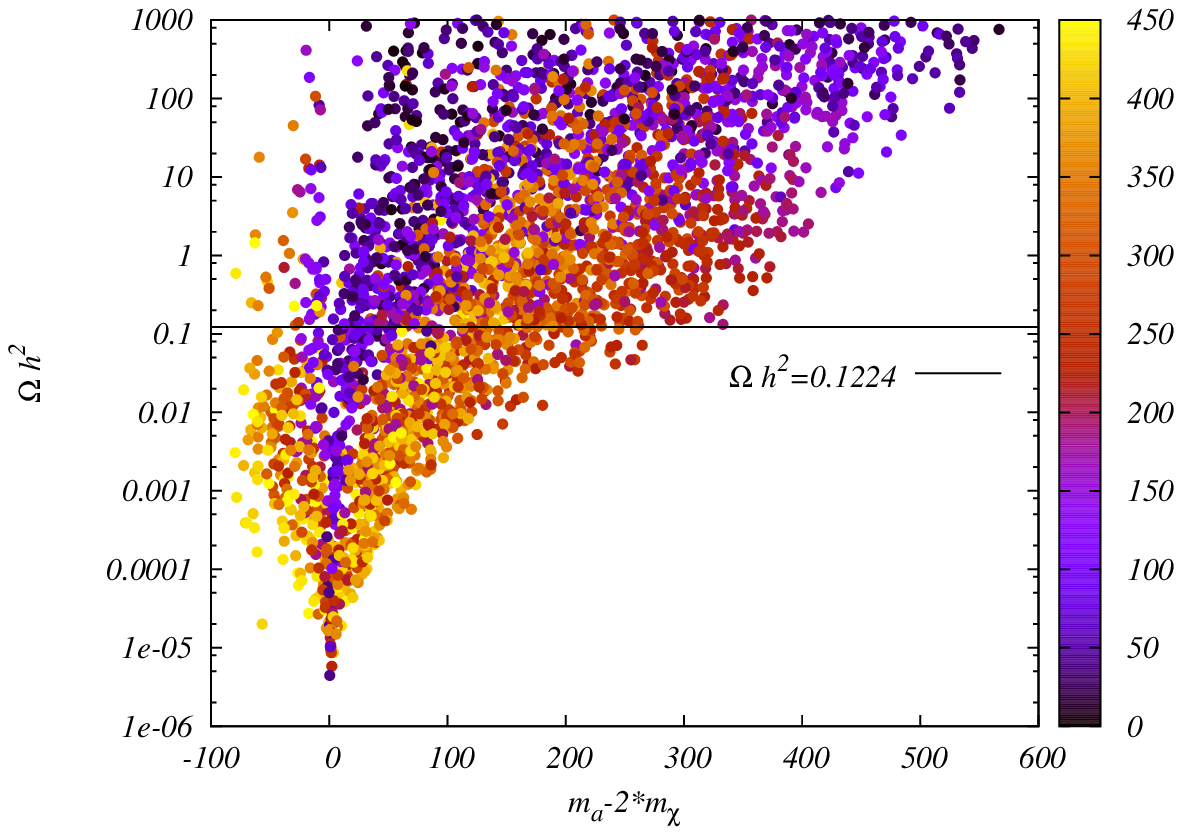}
\includegraphics[width=0.45\textwidth]{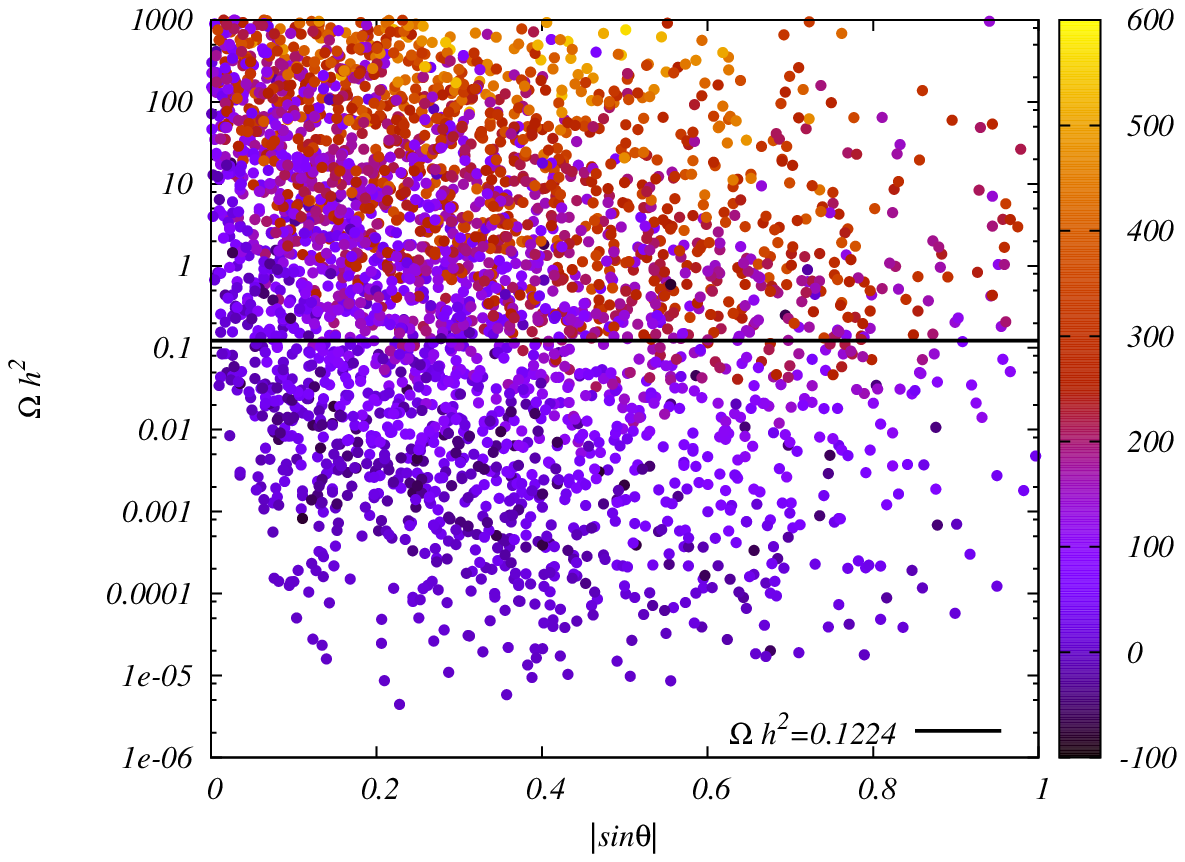}
\caption{\label{fig:om} {\sl Left:} 
Relic density as a function of $m_a-2\,m_\chi$, with the color coding referring to the mass of the DM candidate. {\sl Right:} Relic density as a function of $|\sin\theta|$, with the color coding referring to the mass difference $m_a-2\,m_\chi$. }
\end{center}
\end{figure}
\end{center}
\end{itemize}
All other parameters still populate the original regions (see eqn. (\ref{eq:ranges})).

\section{Predictions for $e^+e^-$ colliders}\label{sec:e+e-}
\begin{figure}[tbh]
\begin{center}
\includegraphics[width=0.45\textwidth]{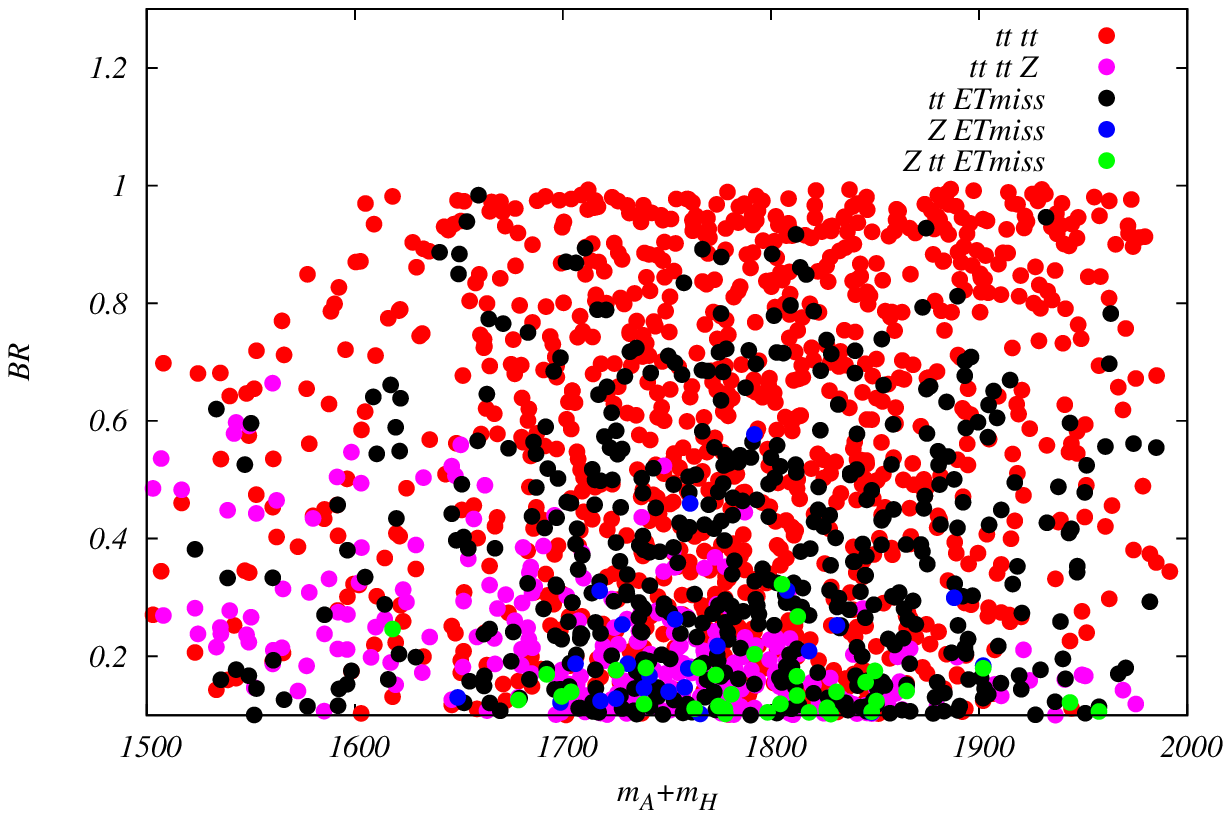}
\includegraphics[width=0.45\textwidth]{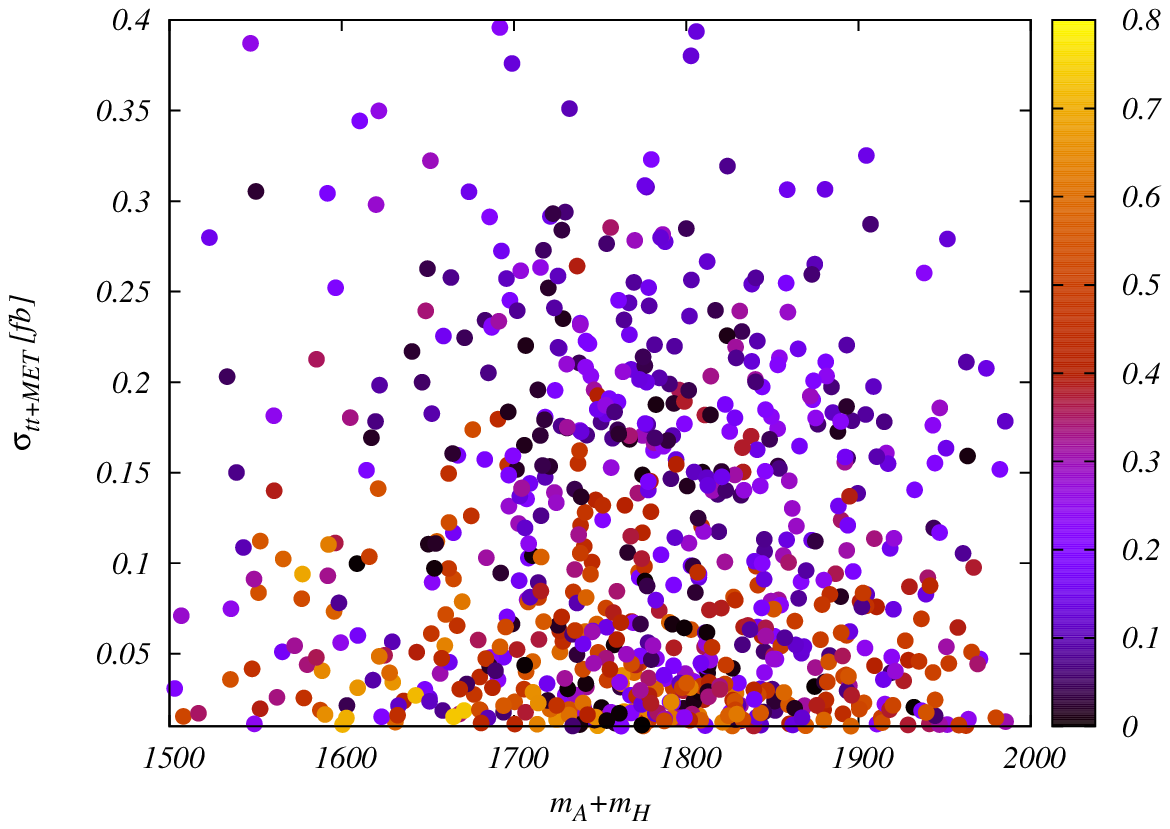}
\caption{\label{fig:brsAH} {\sl Left:} Combined branching ratios for $HA$ final states, as a function of the mass sum. {\sl Right:} Predictions for $tt+\slashed{E}$ rates at a 3 \TeV~ collider, as a function of the mass sum. Color coding refers to the $t\,\bar{t}t\,\bar{t}$ production cross section.}
\end{center}
\end{figure}

\begin{figure}[tbh]
\includegraphics[width=0.49\textwidth]{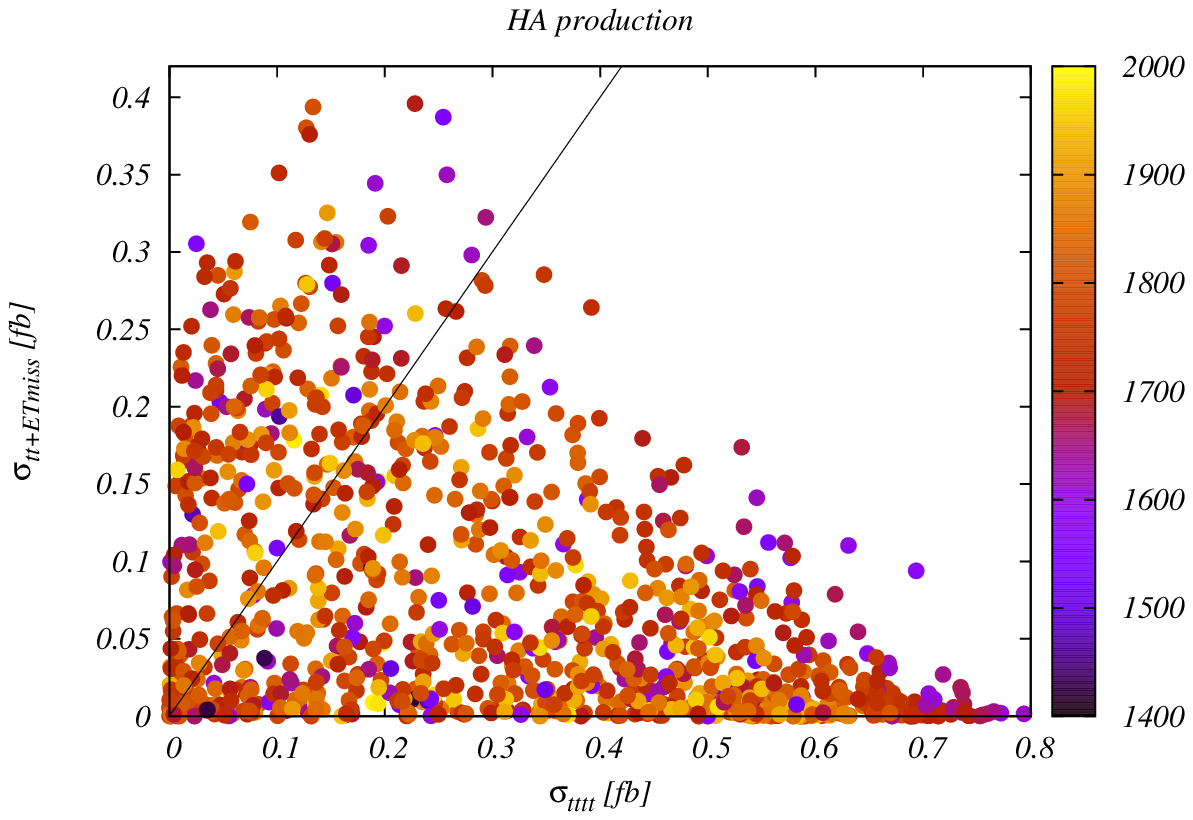}
\includegraphics[width=0.49\textwidth]{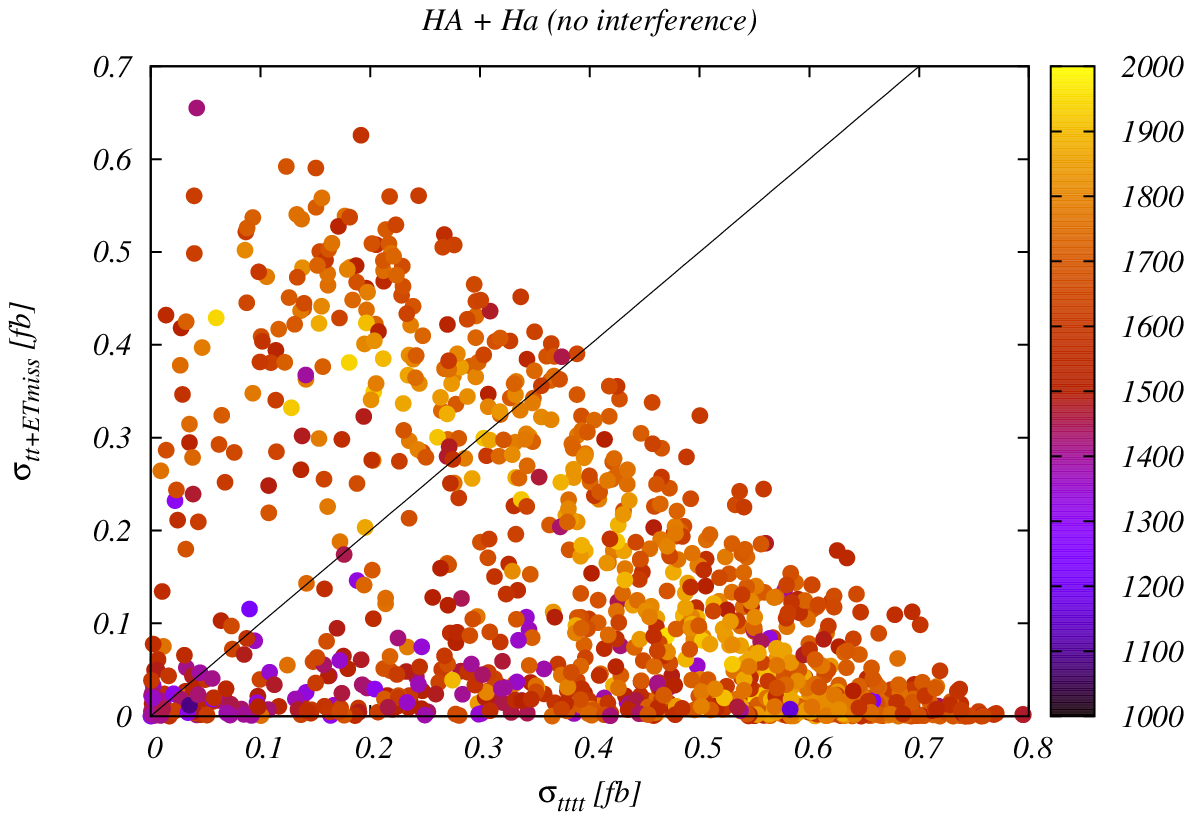}
\caption{\label{fig:xsecs} Production cross sections for $t\bar{t}t\bar{t}$ and $t\bar{t}+\slashed{E}$ final states, using a factorized approach. {\sl (Left:)} via $HA$ and {\sl (right:)} via $HA+Ha$ production. Color coding refers to the mass scale, which is defined as $m_H+m_A$ {\sl (left)}/  $m_H+0.5\,\times\lb m_A+m_a \rb$ {\sl (right)}, respectively.}
\end{figure}

We now show results for rate predictions at $e^+e^-$ colliders. In the limit where $\sin\theta\,\rightarrow\,0$, we recover the decoupling scenario of a standard THDM. It is therefore interesting whether we can find regions in parameter space where novel signatures, and explicitely final states with missing energy, give the largest rates. For $HA\,(a)$ production, which leads to the largest rates after all constraints discussed above have been applied, at a 3 \TeV~collider, production cross-sections can reach up to $1\,\fb$, where largest cross section values are achieved for $m_A+m_H\,\sim\,1400\,\GeV$. Dominant decay modes as well as cross-section predictions for such points are displayed in figure \ref{fig:brsAH}.
$t\bar{t}t\bar{t}$ final states are dominant in large regions of parameter space. The first decay mode that is novel with respect to standard THDMs is the $t\,\bar{t}+\slashed{E}$ final state. In order to identify regions where this state dominates, we show the expected $t\bar{t}t\bar{t}$ and $t\bar{t}+\slashed{E}$ cross sections in figure \ref{fig:xsecs}, where in the right plot we additionally include contributions mediated via $Ha$ production.
We see that indeed we can identify regions where $t\bar{t}+\slashed{E}$ dominates and renders the largest rates.
\section{Summary and outlook}\label{sec:summ}
\vspace{-1mm}
We presented preliminary results for a scan for the THDMa that lets all 12 free parameters of that model float freely, within ranges that were chosen to optimize scan performance. We have identified regions in parameter space that survive all current theoretical and experimental constraints, and provided a first estimate of possible production cross sections within this model at future $e^+ e^-$ facilities, with a focus on signatures not present in a standard THDM. B-physics observables especially impose a lower limit on the charged mass $\,\sim\,800\,\GeV$ that varies with $\tan\be$. Constraints from electroweak precision observables pose relatively strong constraints on the mass differences in the THDMa scalar sector for novel scalars. Requiring the relic density to lie below the current experimental measurement furthermore poses strong constraints on $|m_a-2\,m_\chi|$. 

\noindent
\section*{Acknowledgements}
The author wants to sincerely thank J. Kalinowski, W. Kotlarski, D. Sokolowska, and A.F. Zarnecki for useful discussions in the beginning of this project, and especially W. Kotlarski for help with the setup of the Sarah/ Spheno interface. Further thanks go to M. Misiak and U. Nierste for discussions regarding bounds from B-physics observables, and M. Goodsell as well as the authors of \cite{Abe:2019wjw} for advice. This research was supported in parts by the National Science Centre, Poland, the HARMONIA project under contract UMO-2015/18/M/ST2/00518 (2016-2021), and the OPUS project under contract UMO-2017/25/B/ST2/00496 (2018-2021).

\bibliography{lit}
\end{document}